\documentclass[reprint,superscriptaddress,amsmath,amssymb,aps,prl,longbibliography]{revtex4-1}

\usepackage{graphicx}
\usepackage{dcolumn}
\usepackage{bm}
\usepackage[pdftex,breaklinks,colorlinks=true,citecolor=blue,linkcolor=blue,urlcolor=blue]{hyperref}
\usepackage{ulem}
\usepackage{tabularx}
\begin{document}

\title{Forming a single molecule by magnetoassociation in an optical tweezer}

\author{Jessie~T.~Zhang} 
\email{jessiezhang@g.harvard.edu}
\affiliation{Department of Physics, Harvard University, Cambridge, Massachusetts 02138, USA}
\affiliation{Department of Chemistry and Chemical Biology, Harvard University, Cambridge, Massachusetts 02138, USA}
\affiliation{Harvard-MIT Center for Ultracold Atoms, Cambridge, Massachusetts 02138, USA}

\author{Yichao~Yu} 
\affiliation{Department of Physics, Harvard University, Cambridge, Massachusetts 02138, USA}
\affiliation{Department of Chemistry and Chemical Biology, Harvard University, Cambridge, Massachusetts 02138, USA}
\affiliation{Harvard-MIT Center for Ultracold Atoms, Cambridge, Massachusetts 02138, USA}

\author{William~B.~Cairncross} 
\affiliation{Department of Physics, Harvard University, Cambridge, Massachusetts 02138, USA}
\affiliation{Department of Chemistry and Chemical Biology, Harvard University, Cambridge, Massachusetts 02138, USA}
\affiliation{Harvard-MIT Center for Ultracold Atoms, Cambridge, Massachusetts 02138, USA}

\author{Kenneth~Wang} 
\affiliation{Department of Physics, Harvard University, Cambridge, Massachusetts 02138, USA}
\affiliation{Department of Chemistry and Chemical Biology, Harvard University, Cambridge, Massachusetts 02138, USA}
\affiliation{Harvard-MIT Center for Ultracold Atoms, Cambridge, Massachusetts 02138, USA}

\author{Lewis~R.~B.~Picard} 
\affiliation{Department of Physics, Harvard University, Cambridge, Massachusetts 02138, USA}
\affiliation{Department of Chemistry and Chemical Biology, Harvard University, Cambridge, Massachusetts 02138, USA}
\affiliation{Harvard-MIT Center for Ultracold Atoms, Cambridge, Massachusetts 02138, USA}

\author{Jonathan~D.~Hood}
\altaffiliation[Present address: ]{Department of Chemistry, Purdue University, West Lafayette, Indianna, 47906}
\affiliation{Department of Chemistry and Chemical Biology, Harvard University, Cambridge, Massachusetts 02138, USA}
\affiliation{Harvard-MIT Center for Ultracold Atoms, Cambridge, Massachusetts 02138, USA}

\author{Yen-Wei~Lin} 
\affiliation{Department of Chemistry and Chemical Biology, Harvard University, Cambridge, Massachusetts 02138, USA}
\affiliation{Harvard-MIT Center for Ultracold Atoms, Cambridge, Massachusetts 02138, USA}

\author{Jeremy~M.~Hutson} 
\affiliation{Joint Quantum Centre Durham-Newcastle, Department of Chemistry, Durham University, Durham, DH1 3LE, United Kingdom}

\author{Kang-Kuen~Ni} 
\email{ni@chemistry.harvard.edu}
\affiliation{Department of Chemistry and Chemical Biology, Harvard University, Cambridge, Massachusetts 02138, USA}
\affiliation{Department of Physics, Harvard University, Cambridge, Massachusetts 02138, USA}
\affiliation{Harvard-MIT Center for Ultracold Atoms, Cambridge, Massachusetts 02138, USA}


\begin{abstract}
We demonstrate the formation of a single NaCs molecule in an optical tweezer by magnetoassociation through an \textit{s}-wave Feshbach resonance at 864.11(5)~G. Starting from single atoms cooled to their motional ground states, we achieve conversion efficiencies of $47(1)\%$, and measure a molecular lifetime of $4.7(7)$~ms. By construction, the single molecules are predominantly~($77(5)\%$) in the center-of-mass motional ground state of the tweezer. Furthermore, we produce a single \textit{p}-wave molecule near 807~G by first preparing one of the atoms with one quantum of motional excitation. Our creation of a single weakly bound molecule in a designated internal state in the motional ground state of an optical tweezer is a crucial step towards coherent control of single molecules in optical tweezer arrays.
\end{abstract}

\maketitle

Ultracold polar molecules, with their tunable long-range interactions and rich internal structures, provide a promising means for quantum simulation of novel phases of matter~\cite{Gorshkov2011b,levinsen2011topological,Baranov2012, Sundar2018,Yao2018} and quantum information processing~\cite{DeMille2002,Andre2006,Ni2018,Hudson2018, Hughes2019}. Many key ingredients of these proposals, such as the dipolar exchange interaction~\cite{Yan2013}, long coherence times of nuclear spin and rotational states~\cite{Park2017,Caldwell2020,SeeSelberg2018}, and information transduction between different molecular degrees of freedom~\cite{Lin2019}, have been demonstrated utilizing molecular gases and ions. To realize the aforementioned applications, coherent control of individual ultracold molecules is needed, at the level of single quantum states in both the internal and motional degrees of freedom. A new generation of molecular experiments thus aims towards systems that are simultaneously scalable and able to provide a high level of control over individual particles. This is being pursued through molecular ions in ion traps~\cite{Wolf2016,Chou2017} and through neutral molecules in optical tweezers, both directly cooled~\cite{Anderegg2019} and assembled from their constituent laser-cooled atoms~\cite{Liu2017,Liu2018,Liu2019,Wang2019,Blackmore2018}.

In the bottom-up approach of molecular assembly, forming a single weakly bound molecule is an important milestone towards creating arrays of rovibrational ground-state molecules with tunable interactions. Previously, a single weakly bound molecule (NaCs a${}^3\Sigma, v=-1$) was created from an atom pair by two-photon Raman transfer, but suffered from rapid photon scattering that subsequently scrambled its internal state~\cite{Liu2019}. Here, we form a weakly bound molecule by magnetoassociation through a Fano-Feshbach resonance (FR), which has been established as a robust technique to bridge bi-alkali atoms to rovibrational ground-state molecules, including in bulk gases~\cite{Danzl2008, Ni2008,  Takekoshi2014, Molony2014,Park2015,Guo2016} and in optical lattices \cite{Lang2008,Chotia2012,Moses2015,Reichsollner2017}. While creating molecules in optical tweezers would provide the additional benefits of flexibility and configurability, the peak intensity of optical tweezers is in general several orders of magnitude higher than in optical lattices at the same trapping frequencies; this presents a potential obstacle to the adiabatic magnetoassociation of atoms into molecules. 

In this Letter, we demonstrate that a single Feshbach molecule can be formed in an optical tweezer by magnetoassociation of individually trapped atoms. The molecule has a lifetime of a few milliseconds, limited by scattering from the trap light. By controlling the motional states of the atoms, we can control both the motional and the rotational state of the molecule produced.

The experiment begins with a single $^{23}\text{Na}$ atom and a single $^{133}\text{Cs}$ atom loaded stochastically from a dual-species magneto-optical trap into separate optical tweezers. The atoms are imaged after loading so that we can postselect on whether both species are initially loaded (two-body) or only one is loaded (one-body). The atoms are then simultaneously cooled to their respective 3D motional ground states by polarization gradient cooling and Raman sideband cooling. Details of the trapping and cooling procedures have been reported previously~\cite{Yu2018, Liu2019}. After the atoms are cooled to their motional ground states, we prepare them in the lowest Zeeman energy level: Na~$|F=1, m_F=1\rangle$ and Cs~$|F=3,m_F=3\rangle$. This choice of hyperfine channel eliminates the possibility of spin-changing inelastic collisions, which could occur in our previous work \cite{Hood2019}, and allows production of stable Feshbach molecules.

Our search for Na-Cs FRs was initially guided by multichannel quantum defect theory, using singlet and triplet scattering lengths previously determined from  interaction shift spectroscopy~\cite{Hood2019}. In the present work, we have developed a coupled-channel model of Na+Cs. Singlet and triplet potential curves were obtained by adjusting the interaction potentials in Ref.~\cite{Docenko2006} to reproduce the binding energy of the least-bound triplet state~\cite{Hood2019} and the position of the \textit{s}-wave resonance described below. The adjustment procedures were similar to those used for K+Cs in Ref.~\cite{Grobner2017}. The resulting potential curves were combined with the spin-spin dipolar interaction and an estimate of the 2nd order spin-orbit coupling to allow calculations on states with nonzero relative angular momentum. Scattering calculations were carried out using the MOLSCAT package \cite{molscat:2019,mbf-github:2020} and bound-state calculations using the BOUND and FIELD packages \cite{bound+field:2019, mbf-github:2020}.

\begin{figure}[h]
\centering
\includegraphics{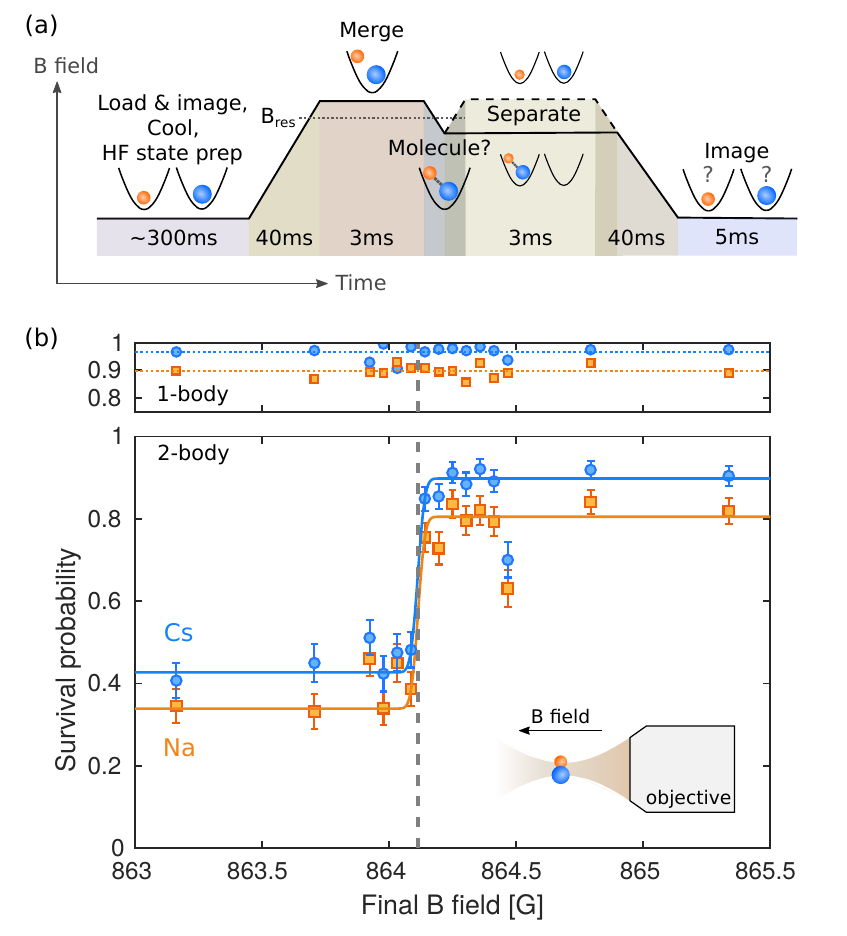}
\caption{\label{fig1} Magnetoassociation at the \textit{s}-wave resonance. (a) Schematic of magnetic field ramp and trap merge and separation sequence as a function of time. Solid (dashed) line indicates one-way molecule conversion (conversion back to atoms for molecule survival detection). Time spent for magnetoassociation is varied for different experiments; see text for details. (b) Determination of the \textit{s}-wave resonance location. The magnetic field is ramped linearly from $866.5$~G to the various magnetic fields at $1$~G/ms. Lower panel: survival probabilities of Na (orange squares) and Cs (blue circles) when both species are loaded. The solid lines are fits to an error function, from which we extract the left and right asymptote values and resonance location. Vertical dashed line indicates resonance location determined from the fit. Upper panel: same experimental run with initial one-body loading. Horizontal dotted lines are the mean values for each species.} 
\end{figure}

In experiments with bulk gases, FRs are usually detected through enhanced 3-body loss~\cite{Chin2010}. The tweezers, however, contain only two atoms, in their lowest-energy states; this precludes 3-body loss and spin-changing inelastic collisions. We therefore directly utilize molecule formation by magnetoassociation for the FR search~\cite{Mark2018}. The experimental sequence for magnetoassociation is shown schematically in Fig.~\ref{fig1}(a). After single atom trapping, ground-state cooling and hyperfine state preparation, a magnetic field produced by a pair of Helmholtz coils is ramped up in $40$~ms along the axial direction of the optical tweezers to $866.5$~G. The two traps are then merged, so that the Na and Cs atoms are held in a single optical tweezer at $1064$~nm and a peak intensity of $81~\text{kW}/\text{cm}^2$, giving trapping frequencies $\omega_\textrm{Cs}=2\pi \times (30,30,5)$~kHz and $\omega_\textrm{Na}\simeq1.07~\omega_\textrm{Cs}$~\cite{Liu2019}. The magnetic field is then ramped linearly down to various values at a rate of $1$~G/ms. If the magnetic field ramp crosses a FR then magnetoassociation is possible. For detection, the tweezer is separated back into the species-specific tweezers before ramping the magnetic field down to zero for imaging the surviving atoms, as shown by the solid line in Fig.~\ref{fig1}(a). Because the imaging detects only the atoms, magnetoassociation to form Feshbach molecules is manifest as a two-body loss.

We locate an \textit{s}-wave Feshbach resonance at $864.11(5)$~G, as shown in the lower panel of Fig.~\ref{fig1}(b); the position is determined by a fit to an error function. An additional loss feature is detected at $864.5$~G, which we attribute to photoassociation enhanced by a narrow resonance nearby \footnote{A d-wave resonance in the \textit{s}-wave scaterring channel is predicted by the coupled-channel model at 866~G. The tweezer power is ramped up to $400$~kW/$\text{cm}^2$ during separation back to individual traps that could lead to FR-enhanced photoassociation.}. As confirmation of the two-body nature of the processes, we also measure the survival rates of the single atoms when loaded without the presence of the other species; these are shown in the upper panel of Fig.~\ref{fig1}(b), and show no features. The contrast between the left and right asymptotes in the two-body loss data gives a molecule conversion efficiency of $47(1)\%$.

\begin{figure}[h]
\centering
\includegraphics{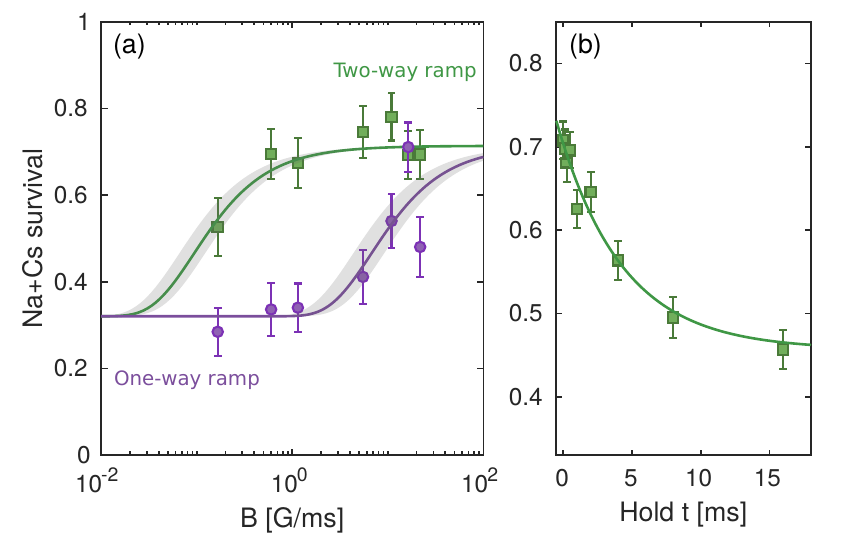}
\caption{\label{fig2} Molecule formation and dissociation efficiencies. (a) Purple circles indicate Na+Cs joint survival probability after magnetoassociation. The magnetic field is ramped linearly from $866$~G to $863.9$~G at different rates. Green squares indicate Na+Cs joint survival probability with an additional reverse magnetic field ramp at the same rate after molecule formation. The solid lines are best-fit curves and the gray shaded areas indicate the errors on the fit. See text for fit details. (b) Lifetime of Feshbach molecule held at $B-B_\text{res}=-0.3$~G and trap intensity $81$~kW/$\text{cm}^2$. Solid line is best fit to an exponential decay.
}
\end{figure}

The atom-to-molecule conversion process can be described by a Landau-Zener (LZ) type avoided crossing with an efficiency that depends on the ramp rate of the magnetic field and characteristic parameters intrinsic to the FR \cite{Thalhammer2006,Chotia2012,Mark2018}. To investigate molecule formation, we vary the rate of a linear magnetic field ramp from $866$~G to $863.9$~G. The resulting joint Na and Cs survival probabilities are shown as the purple circles in Fig.~\ref{fig2}(a). A lower two-body survival probability indicates a higher molecule conversion probability. The one-way molecule conversion efficiency follows the LZ formula $p_\text{mol}=1-e^{-2\pi\delta_\textrm{LZ}}$, where $\delta_\textrm{LZ} = \frac{2\pi n_2}{\mu}\left|\frac{a_\textrm{bg}\Delta}{\dot{B}}\right|$~\cite{Chin2010}. Here $\Delta=1.29$~G and $a_\textrm{bg}=30.7~a_0$ are the width and background scattering length of the Feshbach resonances, obtained from coupled-channel calculations using the method of Ref.~\cite{Frye:2017}, $\mu=19.60$~amu is the reduced mass, $n_2 = \int\int n_\text{Na}(\mathbf{r})\,n_\text{Cs}(\mathbf{r})\,d\mathbf{r}$ is the density of a single pair of Na and Cs atoms in the optical tweezer, and $\dot{B}$ is the magnetic field ramp rate, which is varied experimentally. The purple curve in Fig.~\ref{fig2}(a) is the best fit to the LZ formula. The fit value of the pair density $n_2=2.5(9)\times10^{13}~\text{cm}^{-3}$ is in good agreement with our trap parameters. 

To detect the survival of the Feshbach molecules in the optical tweezer, we dissociate the molecules back into atoms by performing a reverse magnetic field ramp, as shown by the dashed line in Fig.~\ref{fig1}(a). We assume the Feshbach molecule dissociates with certainty since no molecular state exists above resonance \cite{Kohler2006}. The two-way conversion efficiency back to atoms is limited by the time the Feshbach molecules spend in the optical tweezer. This can be expressed as $p_\text{atom}\approx e^{-t_\text{mol}/\bar{\tau}}$, where $t_\text{mol}=\frac{2|B-B_\text{res}|}{\dot{B}}$ is the time spent below the FR, and $\bar{\tau}$ is the molecular lifetime, averaged over the ramped magnetic field. We can directly measure the lifetime of the molecules at a particular magnetic field in a separate experiment by holding the molecules for varying times before dissociating and detecting atom survival. At $B-B_\text{res}=-0.3$~G and a trap peak intensity of 81~kW/cm$^2$, we observe a lifetime of $\tau=4.7(7)$~ms as shown in Fig.~\ref{fig2}(b). The green curve in Fig.~\ref{fig2}(a) is a best fit to the two-way ramp that yields $\tau = 6(2)$~ms, which agrees well with the lifetime measurement and corroborates the lifetime-limited conversion efficiency. This lifetime is promising for future work such as coherently transferring to the rovibrational ground state by stimulated Raman adiabatic passage, which would take tens of microseconds \cite{Ni2008}.

\begin{figure}[h]
\centering
\includegraphics{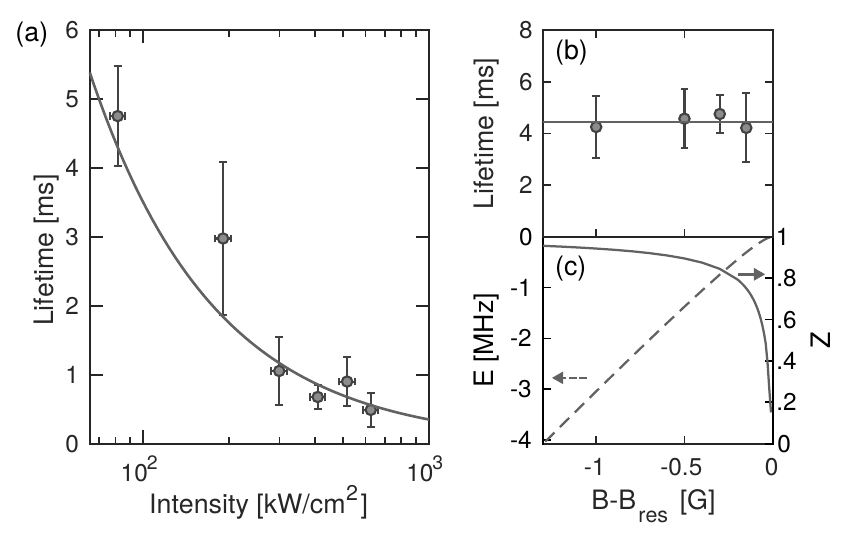}
\caption{\label{fig3} Characterization of \textit{s}-wave Feshbach molecule lifetime.  (a) Dependence on trap intensity. The trap is adiabatically ramped to and held at different intensities after magnetoassociation at $B-B_\text{res}=-0.3$~G. The line is a best fit to inverse scaling. (b) Dependence on magnetic field. The field is ramped to different values during magnetoassociation while the trap intensity is fixed at $81$~kW/$\text{cm}^2$. The solid line is the mean value. (c) The negative binding energy $-E_\textrm{b}$ (dashed line, left axis) and closed-channel fraction Z (solid line, right axis) from coupled-channel bound-state calculations as a function of magnetic field. See text for details.}
\end{figure}

In order to characterize the factors limiting the lifetime of the Feshbach molecules in the optical tweezers, we measure the lifetimes under various hold conditions. In one case, we vary the power of the trap used to hold the molecules after formation. The Feshbach molecules are formed and dissociated with a ramp rate of $3$~G/ms and are held at $B-B_\text{res}=-0.3$~G. We find that the lifetime of the molecules is inversely proportional to trap intensity, as shown in Fig.~\ref{fig3}(a), suggesting that the lifetime is limited by scattering from the trap light. This observation agrees with that previously reported for Feshbach molecules in optical lattices \cite{Chotia2012}. From the scattering rates, we determine the imaginary part of the polarizability at 1064~nm to be 2.8(3)Hz/(kW/cm$^2$); this is $\sim100$ times higher than expected from theoretical predictions~\cite{Dulieu2020}. A similar excessive scattering is also observed in excited NaCs molecular states. There is not yet any clear theoretical explanation of these observations.

We also vary the magnetic field at which the Feshbach molecules are held, over fields that correspond to binding energies up to $E_\textrm{b} = 3$~MHz. The trap peak intensity is fixed at $81$~kW/$\text{cm}^2$. As shown in Fig.~\ref{fig3}(b), we observe no significant variation of the lifetime in this range. The scattering rate of the trapping light depends on the Franck-Condon overlap between the Feshbach molecular state and excited molecular states in the vicinity of the tweezer wavelength; under some circumstances this is proportional to the closed-channel fraction $Z(B)$ of the wavefunction for the Feshbach molecule \cite{Chotia2012}. We have performed coupled-channel bound-state calculations to evaluate $Z(B)$ from the expression $Z(B)=(\mu_\textrm{b}-\mu_\textrm{a}) / (\mu_\textrm{bare}-\mu_\textrm{a})$, where $\mu_\textrm{b}$ ($\mu_\textrm{bare}$) is the magnetic moment of the Feshbach molecular state (the bare molecular state well below threshold), and $\mu_\textrm{a}$ is that of the separated atoms~\cite{Chin2010}. $Z$ and $-E_\textrm{b}$ are shown as functions of magnetic field in Fig.~\ref{fig3}(c). From these we find that this resonance has only a small region of universality. At the magnetic fields we use, $B-B_\text{res}$ between $-1$~G and $-0.15$~G, $Z$ is close to 1 and varies slowly with magnetic field.

The conversion efficiency of an atom pair to a single Feshbach molecule and the motional state of the resulting molecule are both determined by the motional state of the atom pair, described in terms of the relative and center-of-mass (COM) motions~\cite{sm}. Atom pairs can be most efficiently converted to molecules when they are in the ground state of relative motion. We directly measure the population of the relative ground state by interaction shift spectroscopy in a separate experiment, as described in Ref.~\cite{Hood2019}. We find a relative motional ground-state probability of $\sim$58\%, which combined with hyperfine-state preparation fidelities (Na  $\sim$88\%, Cs $\sim$96\%) is consistent with our molecule conversion efficiency of 45-50\%, depending on experimental conditions \cite{sm}. 

The COM motional state of the Feshbach molecule is inherited from that of the constituent atoms. An atom pair that is in its relative motional ground state but an excited COM motional state may still be magnetoassociated to form a molecule. We can infer the atomic COM ground state population from independent Raman sideband thermometry measurements of each atom, and estimate that $77(5)\%$ of the resulting molecules are in the COM motional ground state~\cite{sm}. It should be noted that both the molecule conversion efficiency and the COM ground state population of the Feshbach molecules are not fundamentally limited to their present level, and can be increased by improved atomic ground-state cooling fidelity.

In addition to controlling the motional state of the Feshbach molecule, we can control the internal state through choice of the atomic motional states. In particular, we use a \textit{p}-wave resonance to form a rotationally excited molecule. While atoms in their motional ground states have no relative angular momentum, we can controllably excite the radial motional state of the Na atom by one motional quantum so that the relative motional state of the pair is $\sim24\%$ in the excited state~\cite{sm}. Since the excitation is in the plane perpendicular to the magnetic field axis, as shown in the inset of Fig.~\ref{fig1}(b), the resulting state has relative angular momentum $M_L=\pm1$.

\begin{figure}[t]
\centering
\includegraphics{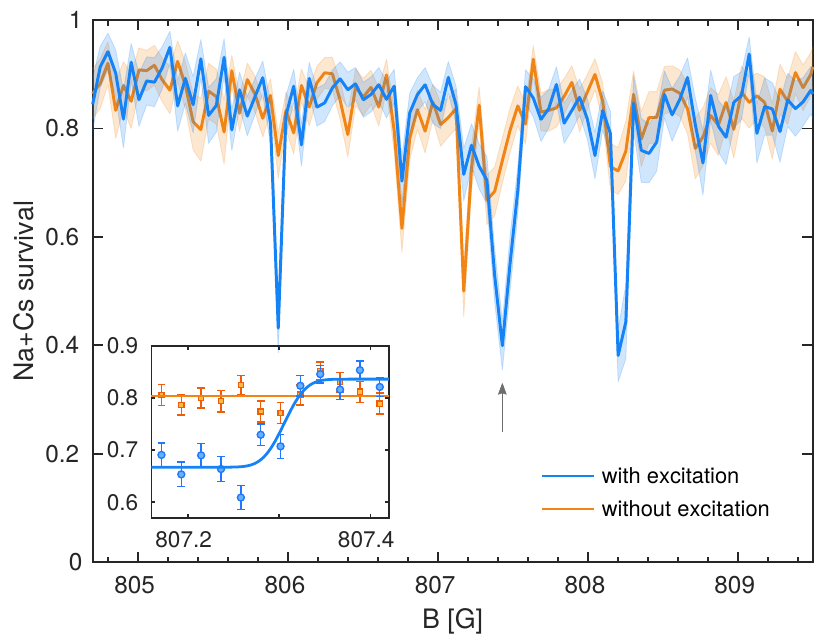}
\caption{\label{fig4} \textit{p}-wave FR spectroscopy by FR-enhanced photoassociation. The atoms are held at a fixed magnetic field in an intense tweezer. Blue (orange) line shows spectrum with (without) motional excitation of Na. Shaded areas indicate error bars. The arrow indicates the resonance used for magnetoassociation in the inset. Inset: \textit{p}-wave Feshbach molecule formation. The magnetic field is ramped linearly from 807.6~G to the various magnetic fields at 0.02~G/ms. Same color scheme as main figure. Blue (orange) curve is fit to error function (mean value across the range).
}
\end{figure}

Our coupled-channel calculations predict two \textit{p}-wave bound states that cross threshold near 807~G, with total molecular spin angular momentum $M_{F,\textrm{b}}=4$ and 5. Each of these splits into components with total angular momentum $M_\textrm{tot}=M_{F,\textrm{b}}$ and $M_{F,\textrm{b}}\pm1$. The colliding atoms have $m_{F,\textrm{Na}}+m_{F,\textrm{Cs}}=4$ and $M_\textrm{tot}=3$ or 5 in the radially excited motional state. We thus expect 3 resonant features for such atoms. We detect these features by FR-enhanced photoassociation \cite{Courteille1998a}, as shown in Fig.~\ref{fig4}. The two atoms are held for $20$~ms in a tweezer with peak intensity 1350~kW/cm$^2$ after merging the traps at a magnetic field value that is scanned. We detect simultaneous two-body loss when the atoms are photoassociated via the excited electronic states by the tweezer light. For comparison, we also show the same scan without the motional excitation on Na. 

As for the \textit{s}-wave Feshbach molecules, we ramp the magnetic field across a \textit{p}-wave Feshbach resonance to transfer the atoms into a \textit{p}-wave molecule. The inset of Fig.~\ref{fig4} shows the survival probability when the magnetic field is ramped linearly down from 807.6~G to various fields at a rate of $0.02$~G/ms in a tweezer held at $81~\text{kW}/\text{cm}^2$ peak intensity. We observe a clear two-body loss feature when we perform the motional excitation, in contrast to the case of no motional excitation. We attribute this to \textit{p}-wave molecule formation and find a conversion efficiency of 16(2)\%.

In conclusion, we have formed single NaCs Feshbach molecules by magnetoassociation in an optical tweezer, using newly identified FRs. In particular, we have formed \textit{s}-wave Feshbach molecules in their motional ground state starting from atoms cooled to their motional ground state and \textit{p}-wave molecules from atoms prepared in specific excited motional states. Feshbach molecules are not susceptible to 3-body collisional losses in optical tweezers, as they are in bulk gases, allowing us to achieve high conversion efficiencies that are not fundamentally limited. While the lifetimes of the Feshbach molecules are limited by scattering from the tweezer light, this does not pose an obstacle to further transfer to the rovibrational ground state by stimulated adiabatic Raman passage. With in-situ atomic rearrangement possible in tweezer arrays \cite{Barredo2016,Endres2016}, the molecule conversion efficiency we have achieved would scale up to a lattice filling fraction near 50\%; this is higher than previously achieved \cite{Moses2015,Reichsollner2017}, and would allow studies of percolating many-body dynamics in 2D geometries \cite{Stauffer}. Identification of ground-state molecules for trap rearrangement might also be possible by imaging the atoms that are not converted to molecules. The high conversion efficiency and exquisite control over the molecules demonstrated here, combined with the configurability provided by the tweezers, render optical tweezer arrays of dipolar molecules a promising platform for quantum simulation and quantum information processing.

\begin{acknowledgments}
We would like to thank Bo Gao for providing the MQDT calculations that guided the initial search for FRs, and Eliot Fenton for experimental assistance. This work is supported by the NSF~(PHY-1806595), the AFOSR~(FA9550-19-1-0089), and the Camille and Henry Dreyfus Foundation~(TC-18-003). J.~T.~Z. is supported by a National Defense Science and Engineering Graduate Fellowship. W.~C. is supported by a Max Planck-Harvard Research Center for Quantum Optics fellowship. K.~W. is supported by an NSF GRFP fellowship. J.~M.~H. is supported by the U.K. Engineering and Physical Sciences Research Council (EPSRC) Grants No.\ EP/N007085/1, EP/P008275/1 and EP/P01058X/1.
\end{acknowledgments}

\bibliography{master_ref}
\bibliographystyle{apsrev4-2}

\end{document}


\title{Supplemental Material: \\ Forming a single molecule by magnetoassociation in an optical tweezer}

\author{Jessie~T.~Zhang} 
\email{jessiezhang@g.harvard.edu}
\affiliation{Department of Physics, Harvard University, Cambridge, Massachusetts 02138, USA}
\affiliation{Department of Chemistry and Chemical Biology, Harvard University, Cambridge, Massachusetts 02138, USA}
\affiliation{Harvard-MIT Center for Ultracold Atoms, Cambridge, Massachusetts 02138, USA}

\author{Yichao~Yu} 
\affiliation{Department of Physics, Harvard University, Cambridge, Massachusetts 02138, USA}
\affiliation{Department of Chemistry and Chemical Biology, Harvard University, Cambridge, Massachusetts 02138, USA}
\affiliation{Harvard-MIT Center for Ultracold Atoms, Cambridge, Massachusetts 02138, USA}

\author{William~B.~Cairncross} 
\affiliation{Department of Physics, Harvard University, Cambridge, Massachusetts 02138, USA}
\affiliation{Department of Chemistry and Chemical Biology, Harvard University, Cambridge, Massachusetts 02138, USA}
\affiliation{Harvard-MIT Center for Ultracold Atoms, Cambridge, Massachusetts 02138, USA}

\author{Kenneth~Wang} 
\affiliation{Department of Physics, Harvard University, Cambridge, Massachusetts 02138, USA}
\affiliation{Department of Chemistry and Chemical Biology, Harvard University, Cambridge, Massachusetts 02138, USA}
\affiliation{Harvard-MIT Center for Ultracold Atoms, Cambridge, Massachusetts 02138, USA}

\author{Lewis~R.~B.~Picard} 
\affiliation{Department of Physics, Harvard University, Cambridge, Massachusetts 02138, USA}
\affiliation{Department of Chemistry and Chemical Biology, Harvard University, Cambridge, Massachusetts 02138, USA}
\affiliation{Harvard-MIT Center for Ultracold Atoms, Cambridge, Massachusetts 02138, USA}

\author{Jonathan~D.~Hood}
\altaffiliation[Present address: ]{Department of Chemistry, Purdue University, West Lafayette, Indiana, 47906}
\affiliation{Department of Chemistry and Chemical Biology, Harvard University, Cambridge, Massachusetts 02138, USA}
\affiliation{Harvard-MIT Center for Ultracold Atoms, Cambridge, Massachusetts 02138, USA}

\author{Yen-Wei~Lin} 
\affiliation{Department of Chemistry and Chemical Biology, Harvard University, Cambridge, Massachusetts 02138, USA}
\affiliation{Harvard-MIT Center for Ultracold Atoms, Cambridge, Massachusetts 02138, USA}

\author{Jeremy~M.~Hutson} 
\affiliation{Joint Quantum Centre Durham-Newcastle, Department of Chemistry, Durham University, Durham, DH1 3LE, United Kingdom}

\author{Kang-Kuen~Ni} 
\email{ni@chemistry.harvard.edu}
\affiliation{Department of Chemistry and Chemical Biology, Harvard University, Cambridge, Massachusetts 02138, USA}
\affiliation{Department of Physics, Harvard University, Cambridge, Massachusetts 02138, USA}
\affiliation{Harvard-MIT Center for Ultracold Atoms, Cambridge, Massachusetts 02138, USA}



\maketitle

\onecolumngrid

\bibliographystyle{apsrev4-2}
\renewcommand*{\citenumfont}[1]{S#1}
\renewcommand*{\bibnumfmt}[1]{[S#1]}

\setcounter{table}{0}
\renewcommand{\thetable}{S\arabic{table}}%
\setcounter{figure}{0}
\renewcommand{\thefigure}{S\arabic{figure}}%
\renewcommand{\thepage}{S\arabic{page}}  
\renewcommand{\thesection}{S\arabic{section}}   
\renewcommand{\theequation}{S.\arabic{equation}}

\newcommand{\na}{\text{Na}}
\newcommand{\cs}{\text{Cs}}
\newcommand\com{\text{com}}
\newcommand\rel{\text{rel}}
\newcommand\acom{a_\text{com}}
\newcommand\arel{a_\text{rel}}
\newcommand\nbar{\bar{n}}

\section{Relative and Center-of-mass motion of atom pair}

The Hamiltonian for a pair of Na and Cs atoms in a 1D harmonic trap, neglecting interactions, can be expressed in the atomic position coordinates as
\begin{align}\label{eq-Hnumbase}
    H &= \sum_{i=\text{Na},\text{Cs}}\frac{p_i^2}{2m_i}+\frac{1}{2}m_i\omega_i^2x_i^2 = \sum_{i=\na,\cs}\hbar\omega_{i}\left(a_i^\dagger a_i+\frac{1}{2}\right).
\end{align}
In particular, $a_i(a_i^\dagger)$ are the annihilation and creation operators for the harmonic modes of the individual atoms in the harmonic trap, which gives rise to a number basis $|n_\na, n_\cs\rangle$ for the motional states. When forming molecules from atom pairs, however, it is more natural to work in terms of the center-of-mass (COM) and relative coordinates, which casts the two-body problem in the lab frame into a one-body problem in the molecular frame of reference. The coordinate transformation is given by
\begin{align}
 x_\text{com}&=\frac{m_\na x_\na+m_\cs x_\cs}{M} \\
 x_\text{rel}&=x_\na-x_\cs
 \end{align}
 and
 \begin{align}
     p_\text{com} &= M\dot x_\com = p_1+p_2 \\ p_\text{rel} &=\mu \dot x_\rel = \frac{m_2p_1-m_1p_2}{M},
\end{align}
 where $M=m_\na+m_\cs$ and $\mu = \frac{m_\na m_\cs}{M}$.
Then the Hamiltonian in eq.~\ref{eq-Hnumbase} can be equivalently expressed as
\begin{align}\label{eq-Hrelbase}
    H= \frac{p_\text{com}^2}{2M}+\frac{1}{2}M\frac{m_\na\omega_\na^2+m_\cs\omega_\cs^2}{M}x_\text{com}^2+\frac{p_\text{rel}^2}{2\mu}+\frac{1}{2}\mu\frac{m_\cs\omega_\na^2+m_\na\omega_\cs^2}{M}x_\text{rel}^2+\mu(\omega_\na^2-\omega_\cs^2)x_\text{com}x_\text{rel}.
\end{align}

The trapping frequencies for the Na and Cs atoms in the optical tweezers $\omega_\na\approx1.07\omega_\cs$ are approximately equal, so we omit the coupling term and take $\omega=\omega_\na=\omega_\cs$ to simplify eq.~\ref{eq-Hrelbase} to
\begin{equation}\label{eq-Harelbase}
    H = \frac{p_\text{com}^2}{2M}+\frac{1}{2}M\omega^2x_\com^2 + \frac{p_\text{rel}^2}{2\mu} +\frac{1}{2}\mu\omega^2x_\rel^2 = \hbar\omega\left(\acom^\dagger\acom+\frac{1}{2}\right)+\hbar\omega\left(\arel^\dagger\arel+\frac{1}{2}\right),
\end{equation}
where 
\begin{align}
\acom^\dagger &= \sqrt{\frac{M\omega}{2\hbar}}\hat x_\com-\frac{i}{\sqrt{2M\omega\hbar}}\hat p_\com \\ \arel^\dagger&=\sqrt{\frac{\mu\omega}{2\hbar}}\hat x_\rel-\frac{i}{\sqrt{2\mu\omega\hbar}}\hat p_\rel
\end{align}
are defined similarly to the atomic case and are related to $a_\na^\dagger,a_\cs^\dagger$ by
\begin{equation}\label{eq-creation_op}
    a_\text{Na}^\dagger = \sqrt{\frac{m_\cs}{M}}a_\text{rel}^\dagger+\sqrt{\frac{m_\na}{M}}a_\text{com}^\dagger,\ \ a_\cs^\dagger = -\sqrt{\frac{m_\na}{M}}a_\text{rel}^\dagger+\sqrt{\frac{m_\cs}{M}}a_\text{com}^\dagger.
\end{equation}
The motional states can then be expressed in terms of a number basis in the COM and relative motional modes $|n_\text{com},n_\text{rel}\rangle$. Our discussion of motional state population pertaining to molecule formation is in terms of this basis.

\subsection{Molecule center-of-mass motional state population}

In the main text we discuss the motional state of the s-wave Feshbach molecules that are formed in the optical tweezers. Due to the adiabatic nature of the magnetoassication process, the motional states of the molecules are inherited from the motional states of the atoms. We can therefore estimate the ground-state population by estimating the COM ground-state population of the constituent atoms. While pairs of atoms must be in the relative motional ground state to be most efficiently magnetoassociated to form molecules, they may be in an arbitrary COM motional state. We use Raman sideband thermometry (RST) on the Na and Cs single atoms~\cite{Yu2018,Liu2019} to infer the portion of the atom pairs capable of forming molecules ($|n_\text{rel}=0\rangle$) that are also in the COM motional ground state ($|n_\text{com}=0, n_\text{rel}=0\rangle$).

Using RST, we can obtain the mean occupation number $\bar{n}_\na,~\bar{n}_\cs$ of each atom along each axis individually. The relative motional ground state population $P(n_\text{rel}=0)$ can be expressed analytically in terms of the $\bar{n}_\na,~\bar{n}_\cs$ values that are measured. In particular, we are interested in finding 
\begin{align}
    P(n_\text{rel}=0)&=\sum_{n=0}^\infty P(n_\text{rel}=0,n_\text{com}=n) \\ &=\sum_{n_1,n_2=0}^\infty P(n_\text{Na}=n_1,n_\text{Cs}=n_2)\left|\langle n_\text{rel}=0,n_\text{com}=n_1+n_2|n_\text{Na}=n_1,n_\text{Cs}=n_2\rangle\right|^2 \label{eq-n1n2} \\
    &= \sum_{n_1,n_2=0}^\infty P(n_\text{Na}=n_1,n_\text{Cs}=n_2)|\langle n_\text{rel}=0,n_\text{com}=n_1+n_2|\frac{(a^\dagger_\na)^{n_1}(a^\dagger_\cs)^{n_2}}{\sqrt{n_1!n_2!}}|n_\text{Na}=0,n_\text{Cs}=0\rangle|^2. \label{eq-n1n2mat}
\end{align}
Eq.~\ref{eq-n1n2} follows from eq.~\ref{eq-Harelbase} since conservation of energy ensures the total excitation numbers in the atomic harmonic modes and COM and relative motional modes must be the same.
The matrix elements in eq.~\ref{eq-n1n2mat} can be found readily using the relations~\ref{eq-creation_op} and noting that we need to consider only terms with $(n_1+n_2)$ COM modal excitations.
\begin{align}
    |\langle n_\text{rel}=&0,n_\text{com}=n_1+n_2|\frac{(a^\dagger_\na)^{n_1}(a^\dagger_\cs)^{n_2}}{\sqrt{n_1!n_2!}}|n_\text{Na}=0,n_\text{Cs}=0\rangle|^2 \\
    &= \frac{1}{n_1!n_2!}\left|\left\langle n_\text{rel}=0,n_\text{com}=n_1+n_2\right|\left(\sqrt{\frac{m_\na}{M}} a^\dagger_\com\right)^{n_1}\left(\sqrt{\frac{m_\cs}{M}} a^\dagger_\com\right)^{n_2}|n_\text{rel}=0,n_\text{com}=0\rangle\right|^2 \\
    &= \frac{m_\na^{n_1}m_\cs^{n_2}}{M^{n_1+n_2}n_1!n_2!}\left|\langle n_\text{rel}=0,n_\text{com}=n_1+n_2|\left(a^\dagger_\com\right)^{n_1}\left(a^\dagger_\com\right)^{n_2}|n_\text{rel}=0,n_\text{com}=0\rangle\right|^2 \\
    &= \left(\frac{m_\na}{M}\right)^{n_1}\left(\frac{m_\cs}{M}\right)^{n_2}\frac{(n_1+n_2)!}{n_1!n_2!}.
\end{align}
With the assumption that the atoms take on a thermal distribution, we have $P(n) = \frac{1}{Z}e^{-\beta\hbar\omega(n+1/2)}$ and $\bar{n} = \frac{1}{e^{\beta\hbar\omega}-1}$, where $\beta=1/k_BT$ and $Z=\frac{e^{-\beta\hbar\omega/2}}{1-e^{-\beta\hbar\omega}}$ is the partition function. Let $\alpha=e^{-\beta\hbar\omega}=\frac{\nbar}{\nbar+1}$, then $P(n) = P(0)\alpha^n$. Using this, we can perform the sum in eq.~\ref{eq-n1n2mat},
\begin{align}
    P(n_\text{rel}=0) &= \sum_{n_1,n_2=0}^\infty P(n_\text{Na}=n_1,n_\text{Cs}=n_2)\left(\frac{m_\na}{M}\right)^{n_1}\left(\frac{m_\cs}{M}\right)^{n_2}\frac{(n_1+n_2)!}{n_1!n_2!} \\
    &= P(n_\na=0)P(n_\cs=0)\sum_{n_1,n_2=0}^\infty \left(\frac{m_\na \alpha_\na}{M}\right)^{n_1}\left(\frac{m_\cs\alpha_\cs}{M}\right)^{n_2}\frac{(n_1+n_2)!}{n_1!n_2!} \\
    &= P(n_\na=0)P(n_\cs=0)\sum_{n=0}^\infty \left(\frac{m_\na \alpha_\na}{M}+\frac{m_\cs\alpha_\cs}{M}\right)^{n}.
\end{align}
Since we necessarily have $(\frac{m_\na \alpha_\na}{M}+\frac{m_\cs\alpha_\cs}{M})<1$, the last expression can be summed up to yield
\begin{align}
    P(n_\text{rel}=0) &= \frac{P(n_\text{Na}=0)P(n_\text{Cs}=0)}{1-\frac{m_\na}{M}\frac{\bar{n}_\na}{\bar{n}_\na+1}-\frac{m_\cs}{M}\frac{\bar{n}_\cs}{\bar{n}_\cs+1}}. \label{eq-3d}
\end{align}
This is the population of atom pairs that can be magnetoassociated to form molecules. On the other hand, the population of the atom pair in the absolute motional ground state is simply given by the joint probability of the two atoms being in their individual ground states,
\begin{align}
    P(n_\text{rel}=0,n_\text{com}=0)&=P(n_\text{Na}=0)P(n_\text{Cs}=0).\label{eq-6d}
\end{align}

Combining eq.~\ref{eq-3d} and eq.~\ref{eq-6d}, we can find the molecular COM ground state fraction
\begin{align}
    P(n_\text{mol-com}=0) &= \frac{P(n_\text{rel}=0,n_\text{com}=0)}{P(n_\text{rel}=0)} \\
    &= 1-\frac{m_\na}{M}\frac{\bar{n}_\na}{\bar{n}_\na+1}-\frac{m_\cs}{M}\frac{\bar{n}_\cs}{\bar{n}_\cs+1} \label{eq-res}.
\end{align}

From RST measurements, we find the occupation numbers of the single Na, Cs atoms along each of the 3 axes (2 radial + axial) to be $\bar{n}_\text{Na}=\{0.09(3),0.07(2),0.30(6)\}$ and $\bar{n}_\text{Cs}=\{0.04(2),0.08(2),0.10(3)\}$. By applying eq.~\ref{eq-res} to each axis individually and taking the product, we find the molecular COM ground state population to be $P(n_\text{mol-com}=0)=77(5)\%$ as stated in the main text.

\subsection{Relative motional excitation}
In the main text we describe forming p-wave molecules in the tweezers by controllably preparing a pair of atoms with relative angular momentum. In particular, starting from atoms cooled to their motional ground states, $|n_\na=0,n_\cs=0\rangle=|n_\text{rel}=0,n_\text{com}=0\rangle$, we prepare the atoms in $|n_\text{rel}=1,n_\text{com}=0\rangle$ by exciting one quantum of motion of Na along the radial direction. We perform the excitation using the heating motional sideband of a two-photon Raman transition; this simultaneously prepares the hyperfine state of the atom. 

We can find the population in the relative-motion excited state after a single excitation of the Na atom using equations~\ref{eq-creation_op}. In particular, we have $\left|\langle n_\text{rel}=1,n_\text{com}=0|a_\text{Na}^\dagger|n_\text{rel}=0,n_\text{com}=0\rangle\right|^2 = \frac{m_\text{Cs}}{M}\approx0.85$. Of note is that exciting the lighter (heavier) atom would give a larger excitation in the relative (COM) motion. Taking into account the fidelity of the Na motional excitation ($75\%$), Cs hyperfine state preparation ($95\%$), and initial relative ground-state population ($40\%$ for the p-wave molecule experiment), we expect $\sim24\%$ of the pairs to be excited in relative motion along one of the radial directions.

\section{Molecule conversion efficiency}
The successful conversion of an atom pair to a single molecule in an optical tweezer relies on several conditions to be simultaneously satisfied. In this section we list the various steps involved and their respective fidelities, which are also summarized in Table \ref{tab-fidelity}, and discuss potential room for improvement in future work.

1. Loading of single atoms. In the present work, we post-process on the experimental runs where both atoms have been trapped. The atom pair probability is therefore assumed to be limited only by the imaging fidelity \cite{Liu2018}, which are Na $0.9996(1)$ and Cs $0.9983(1)$ respectively. When scaling up to multiple tweezers, the tweezer array can be rearranged in situ to obtain near-unity atom filling \cite{Barredo2016,Endres2016}.

2. Population in relative motional ground state. The Na and Cs atoms are cooled to their ground states in their individual traps using Raman sideband cooling. The two traps are then merged so that the atom pair is in a single trap. We measure the relative motional ground-state population of the two atoms in the trap using interaction shift spectroscopy \cite{Hood2019}; under present conditions we find $0.584(44)$. The details for the technical limitations of cooling and merging are detailed in Ref. \cite{Liu2019}.

3. Atom hyperfine state preparation. After the atoms are cooled to their ground states and before merging the two traps, the atoms are optically pumped to the stretched state Na$|F=2,m_F=2\rangle$, Cs$|F=4,m_F=4\rangle$, then driven individually by Raman $\pi$-pulses to the Na$|F=1,m_F=1\rangle$, Cs$|F=3,m_F=3\rangle$ state. The current fidelities are Na 0.882(24) and  Cs 0.956(13), which is limited by purity of optical pumping and dephasing and decoherence during the $\pi$-pulse. This can also be seen in the difference in survival probabilities between the 1-body case and 2-body case above resonance in Figure 1(b) in the main text - any atom pair not in the desired hyperfine state or stretched state is subject to spin-changing collisions that expel both atoms from the trap \cite{Liu2018}.

4. Adiabaticity of magnetic field ramp. As discussed in the main text, this is dependent on magnetic field ramp rate and follows the Landau-Zener formula. At rates below $1$ G/ms in a trap with trapping frequencies $\omega_\text{Cs}=2\pi\times(30,30,5)$ kHz and $\omega_\text{Na}\simeq1.07\omega_\text{Cs}$, we expect the fidelity to be larger than $0.993(6)$.

\begin{table}[ht!]
    \caption{Step-by-step fidelities in molecule conversion}
    \begin{tabular}{l c}
    \hline \hline
    Experimental condition & Fidelity \\
    \hline \hline
    Loading of single atoms (post-selected) & \\
    \ \ \ \ \ Na  & 0.9996(1) \\
    \ \ \ \ \ Cs  & 0.9983(1)\\
    \hline
    Population in relative motional ground state  & 0.584(44) \\
    \hline
    Atom hyperfine state preparation & \\
    \ \ \ \ \ Na\ $|F=1,m_F=1\rangle$ & 0.882(24)\\
    \ \ \ \ \ Cs\ $|F=3,m_F=3\rangle$ &0.956(13) \\
    \hline
    Adiabaticity of magnetic field ramp  & 0.993(6) \\
    \hline \hline
    Overall expected efficiency & 0.48(6) \\
    \hline \hline
    \end{tabular}
\label{tab-fidelity}
\end{table}

\bibliography{master_ref}